\documentclass[groupedaddress,superscriptaddress,aps,prl,preprintnumbers,twocolumn,10pt,fleqn,nofootinbib,nobibnote,eqsecnum,floatfix,showpacs]{revtex4-1}

\usepackage{graphicx}
\usepackage{bm,amsmath,amssymb}
\usepackage[mathscr]{eucal}
\usepackage{color}
\usepackage{afterpage}

\long\def\comment#1{ }
\newcommand{\eqn}[1]{Eq.~\eqref{#1}}
\newcommand{\beq}{\begin{equation}}
\newcommand{\eeq}{\end{equation}}
\newcommand{\nn}{\nonumber\\}

\newcommand{\rmd}{{\rm d}}

\newcommand{\rmI}{{\rm I}}

\newcommand{\abar}{\bar{\alpha}_s}

\newcommand{\tqq}{\theta_{q\bar q}}
\newcommand{\tf}{t_{\rm f}}

\usepackage{color}
\definecolor{darkgreen}{rgb}{0,0.5,0}
\definecolor{darkblue}{rgb}{0,0,0.7}
\definecolor{darkred}{rgb}{0.5,0,0.0}
\definecolor{darkorange}{rgb}{0.8,0.4,0.0}

\begin{document}

\title{Vacuum-like jet fragmentation in a dense QCD medium}

\author{P.~Caucal}
\email{paul.caucal@ipht.fr}
\affiliation{Institut de Physique Th\'{e}orique, Universit\'{e} Paris-Saclay, CNRS, CEA, F-91191 Gif-sur-Yvette, France}

\author{E.~Iancu}
\email{edmond.iancu@ipht.fr}
\affiliation{Institut de Physique Th\'{e}orique, Universit\'{e} Paris-Saclay, CNRS, CEA, F-91191 Gif-sur-Yvette, France}

\author{A.H.~Mueller}
\email{amh@phys.columbia.edu}
\affiliation{Department of Physics, Columbia University, New York, NY 10027, USA}

\author{G.~Soyez}
\email{gregory.soyez@ipht.fr}
\affiliation{Institut de Physique Th\'{e}orique, Universit\'{e} Paris-Saclay, CNRS, CEA, F-91191 Gif-sur-Yvette, France}

\date{\today}

\begin{abstract}
  We study the fragmentation of a jet propagating in a dense
  quark-gluon plasma.
  Using a leading, double-logarithmic approximation in perturbative
  QCD, we compute for the first time the effects of the medium on the
  vacuum-like emissions.
  We show that, due to the scatterings off the plasma, the in-medium
  parton showers differ from the vacuum ones in two crucial aspects:
  their phase-space is reduced and the first emission outside the
  medium can violate angular ordering.
  We compute the jet fragmentation function and find results in
  qualitative agreement with measurements at the LHC.
\end{abstract}

\pacs{
12.38.Bx, 
12.38.Cy, 
12.38.Mh, 
12.39.St, 
25.75.-q 
}
\maketitle

\paragraph{Introduction.}
One of the main objectives of the experimental programs at RHIC and at
the LHC is the characterisation of the quark-gluon plasma (QGP)
produced in ultrarelativistic heavy ion collisions.
An important class of observables used to study this dense form of QCD
matter refer to the physics of {\em jet quenching}, i.e. the
modifications of the properties of an energetic jet or a hadron due to
its interactions with the surrounding medium.
A main source for such modifications, is the {\em medium-induced
  radiation}, responsible e.g. for the energy lost by the jet at large
angles. Within perturbative QCD, this can be computed using the
BDMPS-Z formalism
\cite{Baier:1996kr,Zakharov:1996fv,Wiedemann:2000za}, recently
generalised to include multiple medium-induced branchings
\cite{Blaizot:2012fh,Blaizot:2013hx}.
On top of that, {\em vacuum-like emissions}, triggered by parton
virtualities, should play an important role in the dynamics of the
jets.
A complete picture in which the two mechanisms for radiation are
simultaneously included from first principles is crucially missing.
Existing heuristic scenarios assume for example that vacuum parton
showers develop first, exactly as in the vacuum, and only then interact with the
medium~\cite{Schenke:2009gb,Casalderrey-Solana:2014bpa}; or that
vacuum-like and medium-induced emissions can be combined with each other,
either through postulated medium-modified splitting
functions~\cite{Wang:2001ifa,Majumder:2009ge,Majumder:2013re,Armesto:2009fj}
or within {\em ad-hoc} Monte-Carlo event
generators~\cite{Zapp:2012ak}.

In this Letter, we study for the first time the effects of the medium
on the vacuum-like parton cascades within controlled approximations in
perturbative QCD. Our main conclusion is the existence of a leading,
double-logarithmic, approximation in which such cascades can be factorized
from the medium-induced emissions.
In that limit, jet branching is governed by the usual vacuum (DGLAP)
splitting functions.
These ``vacuum-like'' showers are modified by the medium in two
essential ways, both appearing at leading twist: a constraint on the
allowed phase-space and the possibility for additional radiation at
large angles.
We will show that the jet fragmentation pattern emerging from this
picture is in qualitative agreement with the one observed at the LHC.

For simplicity, we consider a ``jet'' which starts as a colour-singlet
quark-antiquark antenna with a small opening angle $\tqq \ll 1$,
e.g. produced by the decay of a boosted $W/Z$ boson or a virtual photon.
The quark and the antiquark are assumed to have equal energies:
$E_q=E_{\bar q}\equiv E$. We focus on the double-logarithmic
approximation (DLA), in which the parton cascades are dominated by
soft gluons strongly ordered in both energies and emission
angles. However, our picture should remain true in the single-logarithmic
approximation in which we relax the ordering in energy.

When discussing vacuum-like emissions (VLEs) it is useful to
distinguish between three successive stages in the development of a
parton cascade: \texttt{(i)} emissions occurring fully inside the
medium, \texttt{(ii)} a first emission outside the medium (initiated
by a source created inside the medium), and \texttt{(iii)} emissions
from sources created outside the medium. The distinction between
different types of emissions is controlled by the ratio between the
gluon formation time and the ``medium size'' $L$ (the distance
travelled by the jet through the medium).  Clearly, the emissions in
the third stage follow the same pattern as the genuine parton showers
in the vacuum. So, in what follows we shall mainly focus on the first
two stages.

\texttt{(i)} {\em Emissions inside the medium.} These are the
emissions with formation times smaller than $L$.  We start by showing
that the VLEs can be factorised from the medium-induced emissions of
the BDMPS-Z type (triggered by multiple soft scattering in the
medium). This is simply because VLEs proceed fast enough for their
formation not to be influenced by the medium.

\paragraph{Formation times.} To understand that, we compare the
formation times $\tf$ for the vacuum-like and medium-induced emission
of a gluon of energy $\omega$ at an angle $\theta$. $\tf$ is
determined by the condition that the transverse separation
$\Delta r\sim \theta\tf$ between the gluon and its parent parton at
the time of emission be as large as the gluon transverse wavelength
$2/k_\perp$, with $k_\perp\simeq \omega\theta$ its transverse momentum
w.r.t. its emitter. This argument applies to both
vacuum-like and medium-induced emissions and implies
$\tf\simeq {2}{\omega}/k_\perp^2 \simeq {2}/({\omega\theta^2})$.
Then, gluons emitted inside the medium have a minimum $k_\perp$ set by
the momentum acquired via multiple collisions during formation:
$k_{\rm f}^2\simeq \hat q\tf$, with $\hat q$ the jet quenching
parameter.
This translates into an upper limit
$\tf\lesssim \sqrt{2\omega/\hat q}$ on the formation time, leaving two
possibilities: \texttt{(a)} {\it medium-induced emissions}, for which
$k_\perp\simeq k_{\rm f}$, so the corresponding formation time
saturates the upper limit, and \texttt{(b)} {\it vacuum-like
  emissions} (VLEs), for which $k_\perp\gg k_{\rm f}$, with much
shorter formation times.\footnote{Emissions with large
  $k_\perp\gg k_{\rm f}$ can also arise from single hard scatterings
  off the medium. However, such emissions are rare events and by
  themselves do not give rise to parton cascades.}
A VLE is therefore characterised by
\begin{equation}\label{tfvac} 
 \hspace*{-0.2cm} k_\perp\gg k_{\rm f}
  \ \Leftrightarrow\ 
  \frac{2}{\omega\theta^2}\ll\sqrt{\frac{2\omega}{\hat q}}
  \ \Leftrightarrow\ \
  \omega\gg \Big(\frac{2\hat q}{\theta^4}\Big)^{\!\frac{1}{3}}\!\equiv \omega_{0}(\theta).
\end{equation}
The formation time $\tf$ for a medium-induced emission cannot exceed
the medium size $L$, meaning that the conditions~(\ref{tfvac}) are
only effective for $\omega\le \omega_c\equiv \hat q L^2/2$. Emissions
with larger energies ($\omega\ge \omega_c$) behave exactly as in the
vacuum: their emission angle can be arbitrarily small and their
formation time can be larger than $L$.

The above arguments may seem to imply that all the VLEs with energies
$\omega\le \omega_c$ do necessarily have formation times (much)
smaller than $L$. However, there are also VLEs
which evade the conditions~(\ref{tfvac}) because they are emitted
directly {\em outside} the medium: ${2}/({\omega\theta^2})\gtrsim L$.
We discuss such emissions later.

\paragraph{Colour (de)coherence.}
For emissions by a colour-singlet antenna, even a vacuum-like emission
obeying \eqref{tfvac} could be still affected by the medium, via {\em
  color decoherence}
\cite{MehtarTani:2010ma,MehtarTani:2011tz,CasalderreySolana:2011rz,CasalderreySolana:2012ef}.
In the vacuum, gluon emissions at large angles $\theta\gg \tqq$ are
suppressed by the destructive interferences between the quark and the
antiquark. This argument can be iterated to conclude that successive
emissions in the vacuum are {\em ordered in angles},
$\theta_{i+1} \lesssim\theta_i$, an ordering which becomes {\em
  strong} ($\theta_{i+1} \ll \theta_i$) at DLA (see
e.g.~\cite{Dokshitzer:1991wu}). But an antenna propagating through a
dense quark-gluon plasma can lose its coherence via rescattering off
the medium: the quark and the antiquark suffer independent color
rotations, hence the probability that the antenna remains in a color
singlet state decreases with time.
The two legs of the antenna start behaving like independent
color sources after a time $t\sim t_{\rm coh}$, where $t_{\rm coh}$ is
the {\it (de)coherence time}~\cite{CasalderreySolana:2011rz}
\begin{equation}\label{tdec}
  t_{\rm coh}(\tqq)\,\equiv\,\left(\frac{4}{\hat
      q\tqq^2}\right)^{1/3}\,.
\end{equation}
This time scale becomes comparable to $L$ when
$\tqq\sim\theta_c\equiv 2/\sqrt{\hat q L^3}$, with $\theta_c$ the
emission angle for the hardest medium-induced emission, with energy
$\omega_c=\hat q L^2/2$.

Antennas with smaller opening angles $\tqq\lesssim \theta_c$ cannot
lose their coherence, hence their radiation pattern within the medium
is exactly the same in the vacuum. 
One of our main observations is that antennas with $\tqq\gg\theta_c$
--- the most relevant case in practice ---, for which
$t_{\rm coh}(\tqq)\ll L$ and hence could in principle radiate at large
angles $\theta\gg \tqq$, do {\em not} do so to the order of interest:
they only emit at small angles $\theta\lesssim \tqq$, as for coherent
antennas in the vacuum.  To see this, note that
\begin{equation}\label{exp}
\frac{\tf}{t_{\rm coh}}
\,=\,
\frac{(2\hat q\tqq^2)^{1/3}}{\omega\theta^2}
\,=\,\frac{\omega_{0}(\theta)}{\omega}\,
     \left(\frac{\tqq}{\theta}\right)^{2/3}\,.
\end{equation}
The loss of color coherence may only affect the emissions at
sufficiently large angles, $\theta\gtrsim \tqq$, which overlap with
both sources.
For VLEs satisfying (\ref{tfvac}), this implies
${\tf}\ll{t_{\rm coh}}$ meaning that the antenna is still coherent at
the time of the emission and the would-be large-angle emissions
are killed by the interference.
In fine, only emissions with $\theta\lesssim\tqq$ are allowed whether
or not they occur at times larger than the decoherence time~(\ref{tdec}).

\paragraph{Multiple emissions inside the medium.} 
The leading logarithmic behaviour of in-medium parton showers comes
from cascades which are strongly ordered in energies and angles,
i.e. from cascades with $n$ VLE's satisfying
$\tqq\gg\theta_1\gg\cdots \gg\theta_n\gg\theta_c$ and
$E\gg\omega_1\gg\cdots\gg\omega_n\gg\omega_{0}(\theta_n)$.
First, note that the formation times $t_i=2/(\omega_i\theta_i^2)$ are
strongly increasing from one emission to the next. This has the
important consequence that the cascade is formed in a time $t_n$, much
smaller then $L$.
It also means that the condition~\eqref{tfvac} is satisfied by all the
gluons in the cascade if it is satisfied by the last one.
To validate the above picture, we now show that colour coherence
guarantees the angular ordering and that energy loss can be neglected
(at DLA) during the development of the vacuum-like cascade in the medium.

\paragraph{Angular ordering.} For the sake of convenience, colour
coherence is best discussed in the large $N_c$-limit, where the
emission of a soft gluon by an antenna can be described as the
splitting of the original antenna into two daughter antennas.
For any such antenna, say with opening angle $\theta_i$, one can apply
the same argument about angular ordering as for the original antenna
with angle $\tqq$: VLEs at larger angles $\theta > \theta_i$ are
strongly suppressed because their formation times are smaller that the
decoherence time $t_{\rm coh}(\theta_i)$ of that antenna.

\paragraph{Energy loss during formation.}
We are left with showing that the energy lost via medium-induced
radiation remains negligible during the development of a vacuum-like
cascade.
The hardest medium-induced emission that can occur over the time $t$
has an energy $\omega_c(t)\simeq\hat q t^2/2$ and a probability of
order $\alpha_s$~\cite{Baier:1996kr,Zakharov:1996fv}. For
$t=t_n\equiv 2/(\omega_n\theta^2_n)$
\eqn{tfvac} implies $\omega_c(t_n)\ll\omega_n$, i.e. the maximal
energy loss is small compared to the energy of the {\it softest} gluon
in the cascade. The {\em average} energy loss, of order
$\alpha_s\omega_c(t_n)$, is even smaller.
This argument also shows that, over their formation time, the gluons
from the vacuum-like parton showers do not contribute to the energy loss of
the jet. However, after being created, they act as additional sources for
medium-induced radiation (see below).

\begin{figure}[t] \centerline{
\centerline{\includegraphics[width=0.85\columnwidth]{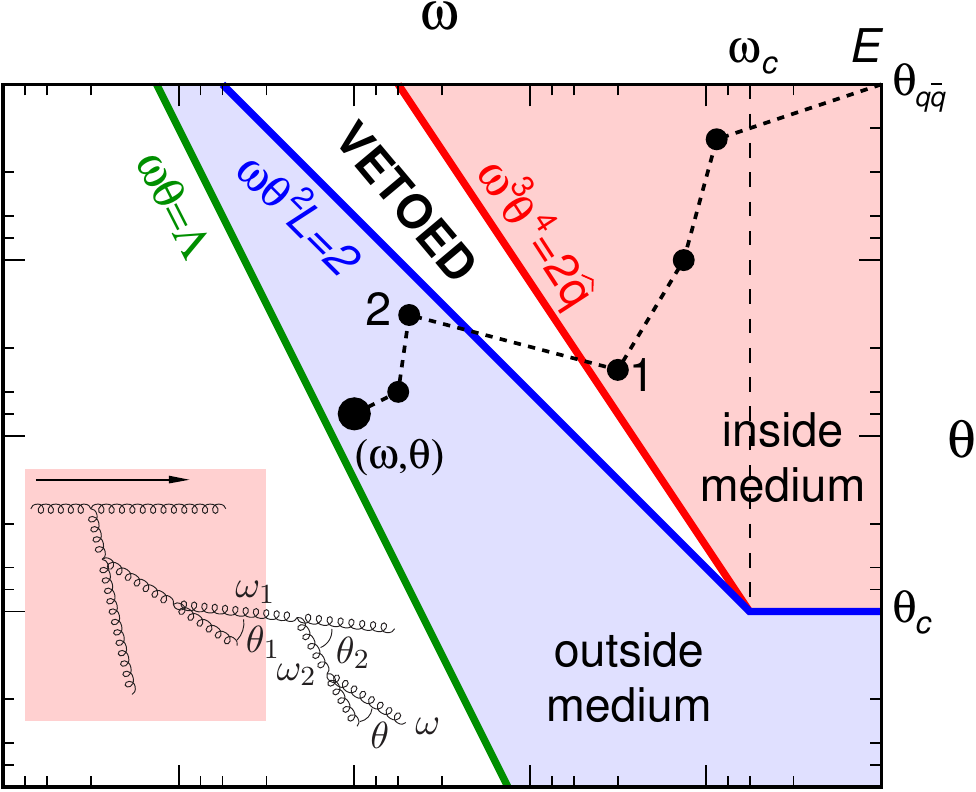}}}
\caption{\small Schematic representation of the phase-space available
  for VLEs, including an example of a cascade with ``1'' the last
  emission inside the medium and ``2'' the first emission outside. }
 \label{phase-space}
\end{figure}

\texttt{(ii)} {\em First emission outside the medium}
The gluons produced inside the medium are not yet on-shell: their
virtualities are as large as their transverse momenta, themselves
bound by the multiple scattering inside the medium:
$k_\perp^2\gg\sqrt{\omega\hat q}\gg\Lambda^2$, with $\Lambda$ the QCD
confinement scale. These partons will thus continue radiating, but
their next VLE must occur {\em outside} the medium, with a large
formation time ${2}/({\omega\theta^2})\gg L$, i.e. with an energy
$\omega\ll \omega_L(\theta)\equiv {2/(L\theta^2)}$. This implies
the existence of a gap in the energy of the VLEs, between the lower
limit $\omega_{0}(\theta)$ on the last gluon emitted {\em inside} the
medium, and the upper limit $\omega_L(\theta)$ on the first gluon
emitted {\em outside} the medium. Since
$\omega_{0}(\theta)=\omega_L(\theta)=\omega_c$ for $\theta=\theta_c$
the gap exists only for $\omega < \omega_c$, as shown in
Fig.~\ref{phase-space}.

\paragraph{No angular ordering.} 
Besides the gap in the phase-space, the medium has another important
effect: the first emission outside the medium can violate angular
ordering. (A similar idea appears in~\cite{Mehtar-Tani:2014yea}.)
Indeed, all the in-medium sources with $\theta\gg\theta_c$ satisfy
$t_{\rm coh}(\theta)\ll L$ and thus lose color coherence after
propagating over a distance $L$ in the medium.
These sources can then radiate at {\em any} angle.\footnote{Notice the
  difference in this respect between in-medium sources emitting inside
  or outside the medium.}
On the contrary, the sources with angles smaller than $\theta_c$
(hence $\omega\gtrsim\omega_c$; see Fig.~\ref{phase-space}), are not
affected by the medium. They behave as if they were created outside
the medium and can radiate only at even smaller angles.

\paragraph{Energy loss after formation.} After being created inside
the medium via VLEs, the partons cross the plasma over a distance of
order $L$ and hence lose energy via medium-induced radiation ---
essentially, as independent colour sources. Whereas this is the main
mechanism for the energy loss by the jet as a whole, it is less
important for the jet fragmentation. Indeed, the {\em typical} gluons
produced via medium-induced radiation
are soft, with 
 $\omega\lesssim\abar^2\omega_c$.  Via successive democratic
branchings \cite{Blaizot:2012fh,Blaizot:2013hx}, they transfer their
energy to many very soft quanta propagating at large angles
$\theta > \tqq$ \cite{Kurkela:2014tla,Blaizot:2014ula,
Iancu:2015uja}.  Hence, such emissions do not
matter for the particle distribution inside the jet.\footnote{One can
  show more rigorously that medium-induced emissions do not matter at
  DLA. However, we believe our physical argument, based on angular
  separation, to be more insightful.}
Furthermore, they do not significantly affect the sources for VLEs:
the energy loss is important only for the sources in a small corner of
the phase-space, at low energies $\omega\lesssim \abar^2\omega_c$ and
large angles, $\theta^2\gtrsim (1/\abar^{3})\theta_c^2$,
cf.~\eqn{tfvac}. We have checked that the effect of
introducing a lower limit $\abar^2\omega_c$ on the energies of the
VLEs is numerically small.
A complete phenomenological picture of jet evolution in the medium
would include medium-induced emissions but, since they go beyond our
current level of approximation, we leave this for future work.

\texttt{(iii)} {\em Emissions from sources created outside the
  medium.}  After a first emission outside the medium, the subsequent
emissions follow, of course, the usual pattern of vacuum-like
cascades, with angular ordering (and energy ordering in our DLA
approximation). The evolution stops when the transverse momentum
$k_\perp\simeq \omega\theta$ becomes comparable to the hadronisation
scale $\Lambda$. This implies a lower boundary,
$\omega\gtrsim \omega_\Lambda(\theta)\equiv \Lambda/\theta$, on the
energy of the produced gluons, shown in Fig.~\ref{phase-space}
together with the other boundaries introduced by the medium.
The most interesting region for gluon production --- the most
sensitive to medium effects highlighted above --- is the ``outside
medium'' region at energies $\omega<\omega_c$.

\paragraph{Gluon distribution.}
Within the present approximation, it is straightforward to compute the
gluon distribution generated by VLEs. 
To that aim we compute the double differential distribution,
\begin{equation}\label{Tdef}
  T(\omega,\theta)\,\equiv\,\omega
  \theta^2\frac{\rmd^2 N}{\rmd\omega\rmd\theta^2}\,,
\end{equation}
which describes the gluon distribution in both energies and emission
angles.  Consider a point with coordinates $(\omega,\,\theta)$ outside
the medium.
A generic contribution to $T(\omega,\theta)$
can be expressed as the product of a vacuum-like cascade inside the
medium, up to an intermediate point $(\omega_1,\,\theta_1)$,
followed by a first emission outside the medium, from
$(\omega_1,\,\theta_1)$ to $(\omega_2,\,\theta_2)$ and, finally, by a
genuine vacuum cascade, from $(\omega_2,\,\theta_2)$ to the measured
point $(\omega,\,\theta)$.  This particular contribution yields (at
large $N_c$)
\begin{align}\label{TP}
&\hspace*{-0.8cm}T(\omega,\theta)=\abar\int_{\theta_{c}^2}^{\tqq^2}\frac{\rmd\theta^2_1}{\theta^2_1}
\int_{\omega_0(\theta_1)}^{E}\frac{\rmd\omega_1}{\omega_1}
\,T_{\rm vac}(\omega_1,\theta_1|E,\tqq)\nn
&\hspace*{-0.6cm}\int^{{\text{min}}(\frac{2}{\omega L},\tqq^2)}_{\theta^2}\frac{\rmd\theta^2_2}{\theta^2_2}
\int_{\omega}^{{\text{min}}(\omega_1,\omega_L\!\,(\theta_2))}\frac{\rmd\omega_2}{\omega_2}\,T_{\rm vac}(\omega,\theta|\omega_2,\theta_2)\,,
 \end{align}
where 
we have chosen $\theta> \theta_c$ for definiteness. The medium effects
enter only via the boundaries of the integrations and no ordering is
assumed between $\theta_1$ and $\theta_2$, in agreement with our
previous discussion.  The explicit factor
$\abar\equiv \alpha_s N_c/\pi$ refers to the first emission outside
the medium and $T_{\rm vac}(\omega_1,\theta_1|E,\tqq)$ represents the
gluon distribution that would be produced in the vacuum to DLA
accuracy, which is well known \cite{Dokshitzer:1991wu}:
\begin{align}\label{TI0}
\hspace*{-0.4cm}T_{\rm vac}(\omega,\theta|E,\tqq)&=
\sum_{n\ge 0}\abar^{n+1}\,\frac{1}{(n!)^2}\,\bigg[\ln\frac{E}{\omega}\,\ln\frac{\tqq^2}{\theta^2}\bigg]^{n}
\nn &=\abar\,\rmI_0\left(2\sqrt{\abar 
\ln\frac{E}{\omega}\,\ln\frac{\tqq^2}{\theta^2}}\right), 
\end{align}
with $\rmI_0(x)$ the modified Bessel function of rank zero. The series
expansion in~(\ref{TI0}) makes explicit the double-logarithmic
enhancement associated with successive emissions
simultaneously ordered in energies and angles.

\begin{figure}[t] \centerline{
\includegraphics[width=0.85\columnwidth]{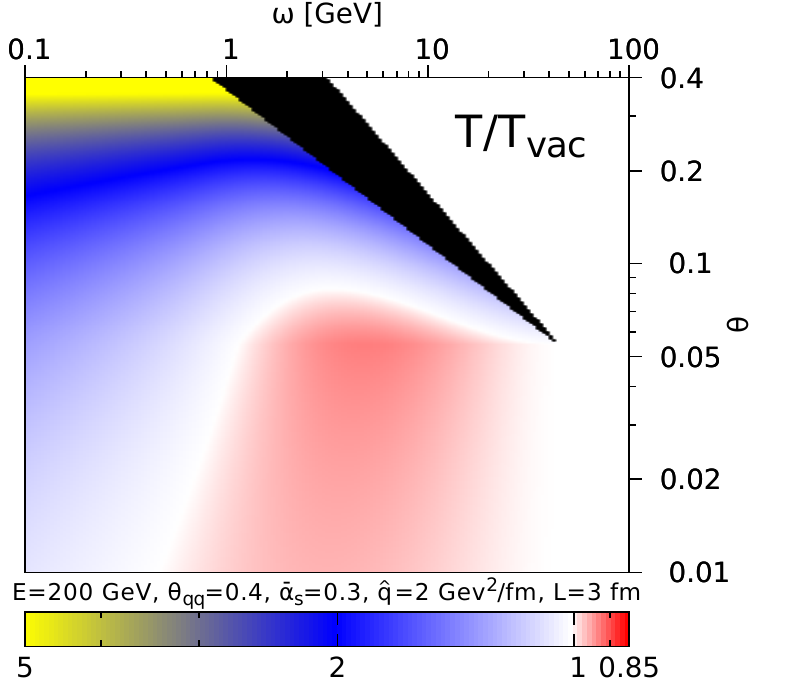}}
 \caption{\small The ratio $T(\omega,\theta^2)/T_{\rm vac}(\omega,\theta^2)$ between the two-dimensional gluon distributions  in the medium and respectively the vacuum, both computed to DLA and for the values of the free parameters $E$, $\tqq$, $\abar$, $\hat q$ and $L$ shown in the figure. }
 \label{result-T}
\end{figure}

Besides \eqref{TP}, there are also contributions in which either the
in-medium cascade, or the out-of-medium one, is missing. For instance,
the first emission by the original antenna can occur directly outside
the medium, in which case there is no cascade inside the medium. 
Our final results include all these contributions.

The result is shown in Fig.~\ref{result-T} where we plot the ratio
$T(\omega,\theta^2)/T_{\rm vac}(\omega,\theta^2)$ for physically
motivated values of the various parameters (specified in the figure).
In our DLA approximation, this ratio is 1 for all the points either
inside the medium or with $\omega >\omega_c$. However, one sees
significant deviations from unity for points outside the medium with
energies $\omega < \omega_c$: for intermediate values of $\omega$ and
relatively small angles $\theta\lesssim 0.1\tqq$, one sees a small but
significative suppression compared to the vacuum (up to 15\%). For
smaller energies and larger angles, $\theta> 0.2$, one rather sees a
strong enhancement, owing to emissions violating angular ordering.

\begin{figure}[t] \centerline{
\includegraphics[width=0.85\columnwidth]{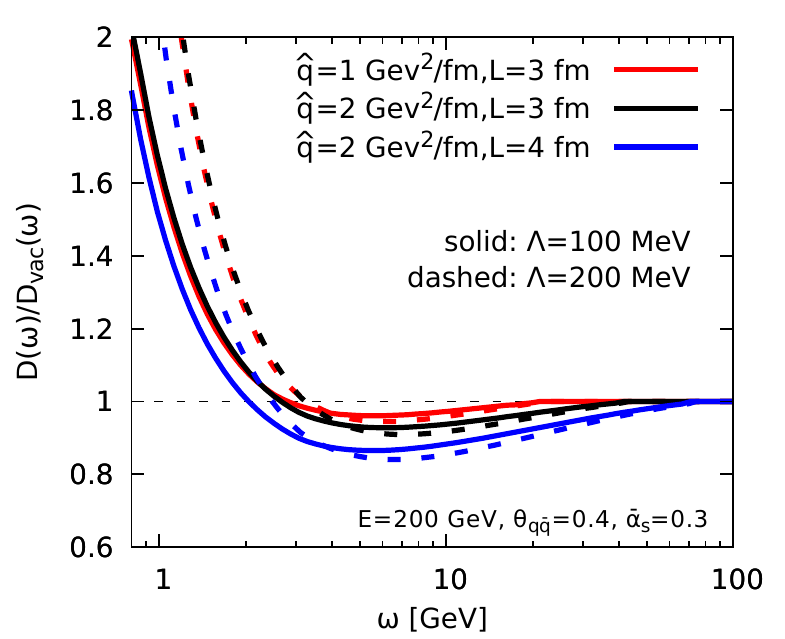}}
 \caption{\small The ratio $D(\omega)/D_{\rm vac}(\omega)$ between the fragmentation functions  in the medium and respectively the vacuum, for different choices for the medium parameters $\hat q$ and $L$ and the hadronisation scale $\Lambda$ (and fixed values for $E$, $\tqq$, and $\abar$).} 
 \label{result-FF}
\end{figure}

\paragraph{The fragmentation function.}
Finally, we use our above results to compute the jet fragmentation
function $D(\omega)$, one of the quantities which is directly measured
in the experiments. This is obtained by integrating the
double-differential distribution over the angles, above the lower
limit $\Lambda/\omega$ introduced by hadronisation: 
\begin{equation}\label{Dx}
D(\omega)\,\equiv\,\omega\frac{\rmd N}{\rmd \omega}
\,=\int_{\Lambda^2/\omega^2}^{\tqq
  ^2}\frac{\rmd\theta^2}{\theta^2}\,T(\omega,\theta^2)\,.
\end{equation}
Our results are shown in Fig.~\ref{result-FF} for several sets of values for the
various parameters. One sees again a slight
suppression (relative to vacuum) at intermediate energies, roughly
from 2~GeV up to an energy $\sim\omega_c$, and a substantial
enhancement at lower energies $\omega \lesssim 2$~GeV. The amount of
suppression at intermediate energies grows with both $\hat q$ and
$L$. This can be easily understood: the logarithmic width of the excluded
region $\ln(\omega_0/\omega_L)=(1/3)\ln(\theta^2/\theta^2_c)$
rises with $1/\theta^2_c=\hat q L^3/4$.  With increasing $L$, the
range in $\omega$ where one sees a enhancement is slowly shrinking, as
expected from the phase-diagram in Fig.~\ref{phase-space}.

Remarkably, the above results are in qualitative agreement with the
LHC measurements of the fragmentation functions for the most central
PbPb collisions~\cite{Chatrchyan:2014ava,Aad:2014wha}. Our picture is
also coherent with jet shape measurements~\cite{Chatrchyan:2013kwa}.
Of course, this should be taken with care, in view of our
approximations. Yet, it supports the idea that our simple picture,
derived in a well-defined limit of perturbative QCD, is close to the
actual physical scenario.
Our picture can be systematically refined to include
e.g. energy conservation and energy loss, and to compute new
observables such as the jet shape.
It can also be implemented as a Monte-Carlo generator.

\smallskip
\noindent{\bf Acknowledgements} We would like to thank Jorge Casalderrey Solana for sharp comments on our manuscript.
The work of E.I. and G.S. is supported in part by the Agence Nationale de la Recherche project 
 ANR-16-CE31-0019-01.   The work of A.H.M.
is supported in part by the U.S. Department of Energy Grant \# DE-FG02-92ER40699. 

\bibliographystyle{utcaps}
\bibliography{refs}

\providecommand{\href}[2]{#2}\begingroup\raggedright\begin{thebibliography}{10}

\bibitem{Baier:1996kr}
R.~Baier, Y.~L. Dokshitzer, A.~H. Mueller, S.~Peigne, and D.~Schiff,
  ``{Radiative Energy Loss of High Energy Quarks and Gluons in a Finite-Volume
  Quark-Gluon Plasma},''
  \href{http://dx.doi.org/10.1016/S0550-3213(96)00553-6}{{\em Nucl. Phys.}
  {\bfseries B483} (1997) 291--320},
\href{http://arxiv.org/abs/hep-ph/9607355}{{\ttfamily arXiv:hep-ph/9607355}}.

\bibitem{Zakharov:1996fv}
B.~G. Zakharov, ``{Fully Quantum Treatment of the Landau-Pomeranchuk-Migdal
  Effect in QED and QCD},'' \href{http://dx.doi.org/10.1134/1.567126}{{\em JETP
  Lett.} {\bfseries 63} (1996) 952--957},
\href{http://arxiv.org/abs/hep-ph/9607440}{{\ttfamily arXiv:hep-ph/9607440}}.

\bibitem{Wiedemann:2000za}
U.~A. Wiedemann, ``{Gluon Radiation Off Hard Quarks in a Nuclear Environment:
  Opacity Expansion},''
  \href{http://dx.doi.org/10.1016/S0550-3213(00)00457-0}{{\em Nucl. Phys.}
  {\bfseries B588} (2000) 303--344},
\href{http://arxiv.org/abs/hep-ph/0005129}{{\ttfamily arXiv:hep-ph/0005129}}.

\bibitem{Blaizot:2012fh}
J.-P. Blaizot, F.~Dominguez, E.~Iancu, and Y.~Mehtar-Tani, ``{Medium-induced
  gluon branching},'' \href{http://dx.doi.org/10.1007/JHEP01(2013)143}{{\em
  JHEP} {\bfseries 1301} (2013) 143},
\href{http://arxiv.org/abs/1209.4585}{{\ttfamily arXiv:1209.4585 [hep-ph]}}.

\bibitem{Blaizot:2013hx}
J.-P. Blaizot, E.~Iancu, and Y.~Mehtar-Tani, ``{Medium-induced QCD cascade:
  democratic branching and wave turbulence},''
  \href{http://dx.doi.org/10.1103/PhysRevLett.111.052001}{{\em Phys.Rev.Lett.}
  {\bfseries 111} (2013) 052001},
\href{http://arxiv.org/abs/1301.6102}{{\ttfamily arXiv:1301.6102 [hep-ph]}}.

\bibitem{Schenke:2009gb}
B.~Schenke, C.~Gale, and S.~Jeon, ``{Martini: an Event Generator for
  Relativistic Heavy-Ion Collisions},''
  \href{http://dx.doi.org/10.1103/PhysRevC.80.054913}{{\em Phys. Rev.}
  {\bfseries C80} (2009) 054913},
\href{http://arxiv.org/abs/0909.2037}{{\ttfamily arXiv:0909.2037 [hep-ph]}}.

\bibitem{Casalderrey-Solana:2014bpa}
J.~Casalderrey-Solana, D.~C. Gulhan, J.~G. Milhano, D.~Pablos, and
  K.~Rajagopal, ``{A Hybrid Strong/Weak Coupling Approach to Jet Quenching},''
  \href{http://dx.doi.org/10.1007/JHEP09(2015)175,
  10.1007/JHEP10(2014)019}{{\em JHEP} {\bfseries 10} (2014) 019},
  \href{http://arxiv.org/abs/1405.3864}{{\ttfamily arXiv:1405.3864 [hep-ph]}}.
[Erratum: JHEP09,175(2015)].

\bibitem{Wang:2001ifa}
X.-N. Wang and X.-f. Guo, ``{Multiple parton scattering in nuclei: Parton
  energy loss},'' \href{http://dx.doi.org/10.1016/S0375-9474(01)01130-7}{{\em
  Nucl. Phys.} {\bfseries A696} (2001) 788--832},
\href{http://arxiv.org/abs/hep-ph/0102230}{{\ttfamily arXiv:hep-ph/0102230
  [hep-ph]}}.

\bibitem{Majumder:2009ge}
A.~Majumder, ``{Hard collinear gluon radiation and multiple scattering in a
  medium},'' \href{http://dx.doi.org/10.1103/PhysRevD.85.014023}{{\em Phys.
  Rev.} {\bfseries D85} (2012) 014023},
\href{http://arxiv.org/abs/0912.2987}{{\ttfamily arXiv:0912.2987 [nucl-th]}}.

\bibitem{Majumder:2013re}
A.~Majumder, ``{Incorporating Space-Time Within Medium-Modified Jet Event
  Generators},'' \href{http://dx.doi.org/10.1103/PhysRevC.88.014909}{{\em Phys.
  Rev.} {\bfseries C88} (2013) 014909},
\href{http://arxiv.org/abs/1301.5323}{{\ttfamily arXiv:1301.5323 [nucl-th]}}.

\bibitem{Armesto:2009fj}
N.~Armesto, L.~Cunqueiro, and C.~A. Salgado, ``{Q-PYTHIA: A Medium-modified
  implementation of final state radiation},''
  \href{http://dx.doi.org/10.1140/epjc/s10052-009-1133-9}{{\em Eur. Phys. J.}
  {\bfseries C63} (2009) 679--690},
\href{http://arxiv.org/abs/0907.1014}{{\ttfamily arXiv:0907.1014 [hep-ph]}}.

\bibitem{Zapp:2012ak}
K.~C. Zapp, F.~Krauss, and U.~A. Wiedemann, ``{A perturbative framework for jet
  quenching},'' \href{http://dx.doi.org/10.1007/JHEP03(2013)080}{{\em JHEP}
  {\bfseries 03} (2013) 080},
\href{http://arxiv.org/abs/1212.1599}{{\ttfamily arXiv:1212.1599 [hep-ph]}}.

\bibitem{MehtarTani:2010ma}
Y.~Mehtar-Tani, C.~A. Salgado, and K.~Tywoniuk, ``{Antiangular Ordering of
  Gluon Radiation in QCD Media},''
  \href{http://dx.doi.org/10.1103/PhysRevLett.106.122002}{{\em Phys. Rev.
  Lett.} {\bfseries 106} (2011) 122002},
\href{http://arxiv.org/abs/1009.2965}{{\ttfamily arXiv:1009.2965 [hep-ph]}}.

\bibitem{MehtarTani:2011tz}
Y.~Mehtar-Tani, C.~A. Salgado, and K.~Tywoniuk, ``{Jets in QCD Media: from
  Color Coherence to Decoherence},''
  \href{http://dx.doi.org/10.1016/j.physletb.2011.12.042}{{\em Phys. Lett.}
  {\bfseries B707} (2012) 156--159},
\href{http://arxiv.org/abs/1102.4317}{{\ttfamily arXiv:1102.4317 [hep-ph]}}.

\bibitem{CasalderreySolana:2011rz}
J.~Casalderrey-Solana and E.~Iancu, ``{Interference Effects in Medium-Induced
  Gluon Radiation},'' \href{http://dx.doi.org/10.1007/JHEP08(2011)015}{{\em
  JHEP} {\bfseries 08} (2011) 015},
\href{http://arxiv.org/abs/1105.1760}{{\ttfamily arXiv:1105.1760 [hep-ph]}}.

\bibitem{CasalderreySolana:2012ef}
J.~Casalderrey-Solana, Y.~Mehtar-Tani, C.~A. Salgado, and K.~Tywoniuk, ``{New
  picture of jet quenching dictated by color coherence},''
  \href{http://dx.doi.org/10.1016/j.physletb.2013.07.046}{{\em Phys.Lett.}
  {\bfseries B725} (2013) 357--360},
\href{http://arxiv.org/abs/1210.7765}{{\ttfamily arXiv:1210.7765 [hep-ph]}}.

\bibitem{Dokshitzer:1991wu}
Y.~L. Dokshitzer, V.~A. Khoze, A.~H. Mueller, and S.~I. Troian, ``{Basics of
  perturbative QCD},''. Gif-sur-Yvette, France. Ed. Frontieres (1991) 274 p.

\bibitem{Mehtar-Tani:2014yea}
Y.~Mehtar-Tani and K.~Tywoniuk, ``{Jet (de)coherence in PbPb collisions at the
  LHC},'' \href{http://dx.doi.org/10.1016/j.physletb.2015.03.041}{{\em Phys.
  Lett.} {\bfseries B744} (2015) 284--287},
\href{http://arxiv.org/abs/1401.8293}{{\ttfamily arXiv:1401.8293 [hep-ph]}}.

\bibitem{Kurkela:2014tla}
A.~Kurkela and U.~A. Wiedemann, ``{Picturing perturbative parton cascades in
  QCD matter},'' \href{http://dx.doi.org/10.1016/j.physletb.2014.11.054}{{\em
  Phys.Lett.} {\bfseries B740} (2015) 172--178},
\href{http://arxiv.org/abs/1407.0293}{{\ttfamily arXiv:1407.0293 [hep-ph]}}.

\bibitem{Blaizot:2014ula}
J.-P. Blaizot, Y.~Mehtar-Tani, and M.~A.~C. Torres, ``{Angular structure of the
  in-medium QCD cascade},''
  \href{http://dx.doi.org/10.1103/PhysRevLett.114.222002}{{\em Phys.Rev.Lett.}
  {\bfseries 114} no.~22, (2015) 222002},
\href{http://arxiv.org/abs/1407.0326}{{\ttfamily arXiv:1407.0326 [hep-ph]}}.

\bibitem{Iancu:2015uja}
E.~Iancu and B.~Wu, ``{Thermalization of mini-jets in a quark-gluon plasma},''
  \href{http://dx.doi.org/10.1007/JHEP10(2015)155}{{\em JHEP} {\bfseries 10}
  (2015) 155},
\href{http://arxiv.org/abs/1506.07871}{{\ttfamily arXiv:1506.07871 [hep-ph]}}.

\bibitem{Chatrchyan:2014ava}
{\bfseries CMS} Collaboration, S.~Chatrchyan {\em et al.}, ``{Measurement of
  jet fragmentation in PbPb and pp collisions at $\sqrt{s_{NN}}=2.76$ TeV},''
  \href{http://dx.doi.org/10.1103/PhysRevC.90.024908}{{\em Phys. Rev.}
  {\bfseries C90} no.~2, (2014) 024908},
\href{http://arxiv.org/abs/1406.0932}{{\ttfamily arXiv:1406.0932 [nucl-ex]}}.

\bibitem{Aad:2014wha}
{\bfseries ATLAS} Collaboration, G.~Aad {\em et al.}, ``{Measurement of
  inclusive jet charged-particle fragmentation functions in Pb+Pb collisions at
  $\sqrt{s_{NN}}$=2.76 TeV with the ATLAS detector},''
  \href{http://dx.doi.org/10.1016/j.physletb.2014.10.065}{{\em Phys. Lett.}
  {\bfseries B739} (2014) 320--342},
\href{http://arxiv.org/abs/1406.2979}{{\ttfamily arXiv:1406.2979 [hep-ex]}}.

\bibitem{Chatrchyan:2013kwa}
{\bfseries CMS Collaboration} Collaboration, S.~Chatrchyan {\em et al.},
  ``{Modification of jet shapes in PbPb collisions at $\sqrt {s_{NN}} = 2.76$
  TeV},'' \href{http://dx.doi.org/10.1016/j.physletb.2014.01.042}{{\em
  Phys.Lett.} {\bfseries B730} (2014) 243--263},
\href{http://arxiv.org/abs/1310.0878}{{\ttfamily arXiv:1310.0878 [nucl-ex]}}.

\end{thebibliography}\endgroup

\end{document}